\documentclass[%
 reprint,
superscriptaddress,
 amsmath,amssymb,
 aps,
]{revtex4-2}

\usepackage{graphicx}% Include figure files
\usepackage{dcolumn}% Align table columns on decimal point
\usepackage{bm}% bold math
\usepackage{xcolor}
\usepackage{physics}
\begin{document}

\preprint{APS/123-QED}

\title{Spin-orbit maximally discordant mixed states}% Force line breaks with \\
%%\thanks{A footnote to the article title}%

\author{D. G. Braga}
\email{danielbraga@id.uff.br}
\affiliation{Instituto de Ci\^encias Exatas, Universidade Federal Fluminense, 27213-145, Volta Redonda, Rio de Janeiro, Brazil}
\affiliation{Instituto de F\'{i}sica, Universidade Federal Fluminense, Av. Gal. Milton Tavares de Souza s/n, Gragoat\'{a}, 24210-346 Niter\'{o}i, Rio de Janeiro, Brazil}

\author{I. Fonseca}
\email{igorrf@id.uff.br}
\affiliation{Instituto de Ci\^encias Exatas, Universidade Federal Fluminense, 27213-145, Volta Redonda, Rio de Janeiro, Brazil}
\affiliation{Instituto de F\'{i}sica, Universidade Federal Fluminense, Av. Gal. Milton Tavares de Souza s/n, Gragoat\'{a}, 24210-346 Niter\'{o}i, Rio de Janeiro, Brazil}

\author{W. F. Balthazar}
\email{wagner.blthzr@gmail.com}
%\affiliation{Instituto de Ci\^encias Exatas, Universidade Federal Fluminense, 27213-145, Volta Redonda, Rio de Janeiro, Brazil}
\affiliation{International Iberian Nanotechnology Laboratory (INL), Av. Mestre José Veiga, 4715-330 Braga, Portugal.}
\affiliation{Instituto Federal do Rio de Janeiro, 27213-100, Volta Redonda, Rio de Janeiro, Brazil}.

\author{M. S. Sarandy}
\email{msarandy@id.uff.br}
\affiliation{Instituto de F\'{i}sica, Universidade Federal Fluminense, Av. Gal. Milton Tavares de Souza s/n, Gragoat\'{a}, 24210-346 Niter\'{o}i, Rio de Janeiro, Brazil}

\author{J. A. O. Huguenin}
\email{jose\_huguenin@id.uff.br}
\affiliation{Instituto de Ci\^encias Exatas, Universidade Federal Fluminense, 27213-145, Volta Redonda, Rio de Janeiro, Brazil}
\affiliation{Instituto de F\'{i}sica, Universidade Federal Fluminense, Av. Gal. Milton Tavares de Souza s/n, Gragoat\'{a}, 24210-346 Niter\'{o}i, Rio de Janeiro, Brazil}

\date{\today}% It is always \today, today,

\begin{abstract}
 We introduce a proposal to prepare spin-obit maximally discordant mixed states by a linear optical circuit, with quantum bits (qubits) encoded in the polarization and transverse mode degrees of freedom of photons. In particular, we discuss how to prepare non-balanced spin-orbit entangled states, applying this technique to obtain maximally discordant mixed states. We present a simulation of the optical circuit by using the Jones matrix formalism. %, including tomographic measurements.
 We performed a study of entanglement, classical and quantum correlations. The results show excellent agreement with the underlying theory and open a new experimental approach for addressing quantum correlations in optical setups.          
\end{abstract}
 
\maketitle

%\tableofcontents

\section{\label{sec:1-intro}Introduction}

%%%SDiscord--
Quantum correlations are essential for the realization of quantum information tasks, being entanglement their main resource~\cite{nielsen00}. In addition to entanglement, other (classical and quantum) correlations may play an important role in quantum protocols, such as quantum discord (QD)~\cite{Zurek, DiscordQuantComp}. QD characterizes quantunmess through the inability of a quantum state to be kept undisturbed by a non-selective measurement. QD has provided tools in several distinct scenarios, e.g. quantum spin models~\cite{DiscordSpinModels} critical quantum systems~\cite{DiscordCritcSys}, quantum metrology~\cite{Modi:11}, environment-induced sudden transitions~\cite{LearningDat-book}, and multiqubit systems~\cite{DiscordMultiqubits}. In photonic systems, QD has been measured in two-photon states, exploring linear optical setups~\cite{twophptonDiscord}. 

In the context of QD, quantum correlated states inherently present classical correlations. The total correlation, as provided by the mutual information, is typically divided into a quantum contribution as well as a classical counterpart. A natural question is then how to maximize the quantum contribution, as measured by QD, for a fixed amount of classical correlation. This maximization leads to the so called maximally discordant mixed state (MDMS)~\cite{GALVEMSDS}. For a pure state, it is well known that a MDMS is also a maximally entangled state, i.e., a Bell state. However, for mixed states, an unbalanced Bell state can be used to construct a MDMS. The generation of MDMS of two qubits in the absence of quantum entanglement has been recently proposed by using two dissipative schemes, which are based on the interaction between four-level atoms and strongly lossy optical cavities~\cite{MDMSdissipation}. 

%%%Spin-orbit--
%In the experimental scenario,
From an architecture point of view, optical systems are reliable sources of entanglement~\cite{ep1}, with qubits easily encoded in the degrees of freedom (DoF) of light. As examples of the use of DoF of light in optical systems to realize quantum information protocols, we can mention the implementation of quantum gates~\cite{qgp1}, quantum computing protocols~\cite{quantcomp}, teleportation~\cite{qtelep}, and quantum cryptography~\cite{qcrypt}. Moreover, an efficient and scalable protocol for quantum computing by using linear optical elements and projective measurements in single-photon qubits has been proposed by Knill, Laflamme and Milburn \cite{KLM}, yielding an intense field of research~\cite{KLMDeteerministic, KLMmeasurebased, RevModPhysKLM}. 
On the other hand, the encoding of qubits in polarization and first-order Hermite-Gaussian (HG) modes has given rise to the so called spin-orbit modes. In turn, the classical-quantum analogy of intense laser beam allowed for the implementation of quantum protocols with linear optical circuits for spin-orbit modes associated with intense laser beams that emulate single-photon experiments. A remarkable example was the topological phase predicted in the evolution of a pair of entangled qubits, which has been realized in this scenario \cite{topo}. Further results include  experiments for bipartite systems showing quantum inequalities violation by spin-orbit maximally non-separable modes, such as Bell's inequality~\cite{bell1, bell3} and contextuality~\cite{li2017experimental, context}. By adding path DoF to polarization and transverse mode, a tripartite system can be emulated and Mermin's inequality can be shown to be violated~\cite{tript}. Such an approach has been successfully employed and demonstrated the power of spin-orbit modes in quantum information tasks, with applications in the experimental study of environment-induced entanglement~\cite{environSO}, quantum cryptography~\cite{ccrypt}, and quantum gates~\cite{cqg1, Braz:14, cqg3}.

Concerning QD, it appears in the scenario of spin-orbit modes in the proposal of an optical circuit to prepare spin-orbit X-states, where QD has been derived for different classes of states~\cite{Xstate}. However, the MDMS has not been explicitly considered. 
In this work we present a proposal to prepare MDMS by using spin-orbit modes in linear optical circuits. We then evaluate QD for the states resulting from the simulation of the circuit. The paper is organized as follows. In section II, we present the theoretical background. We present the MDMS and the QD analysis. In Section III, we present in details the linear optical circuit proposed to the preparation of the MDMS by using spin-orbit modes. Section IV is devoted to presenting the results of numerical simulation of the optical circuit by means Jones formalism. Finally, in Section V, conclusions are discussed.   

\section{\label{sec:2-MDS} QD and MDMS}

Let us begin by reviewing the theoretical approach to compute QD and the definition of the MDMS. 
Here, we used the definition of entropic QD, following the steps developed in Refs.~\cite{Luo, Girolami2011}. 
In classical information theory, the uncertainty about a random variable $A$, which can assume the values $a$ with corresponding probability $p_{a}$, is provided by the Shannon's entropy $H(A)=-\sum_{a}p_{a}\log_{2}p_{a}$. The joint uncertainty about two random variables $A$ and $B$ is $H(AB)=-\sum_{a,b}p_{a,b}\log_{2}p_{a,b}$, with $p_{a,b}$ being the joint probability distribution. The total amount of correlation between $A$ and $B$ is given by the difference in the uncertainty about $A$ before and after the variable $B$ is known, namely, $\mathcal{J}(A:B)=H(A)-H(A|B)$, where $H(A|B)=-\sum_{a,b}p_{a,b}\log_{2}p_{a|b}$ is the conditional entropy, with $p_{a|b}$ the probability for $A=a$ given that $B=b$. From Bayes rule, $p_{a|b}=p_{a,b}/p_{b}$. Then, we can rewrite $\mathcal{J}(A:B)$ in the equivalent
form $\mathcal{I}(A:B)=H(A)+H(B)-H(AB)$.

For the quantum case, we consider a bipartite system $AB$ described by a composite density operator $\rho$. Then, the total amount of correlation between the two subsystems $A$ and $B$ described by local density operators $\rho_A$ and $\rho_B$, respectively, is given by the quantum mutual information
\begin{equation}
        I_m(\rho)=S(\rho_A)+S(\rho_B)-S(\rho),
\end{equation}
where $S(\rho)=-Tr[\rho \log_2 \rho]$ is the von Neumann entropy. We can also generalize $\mathcal{J}(A:B)$ to the quantum realm by considering a measurement over subsystem $B$, with measurement operators denoted by $\{B_k\}$. 
The composite state $\rho$ then collapses to $\rho_k$ with probability $p_k$. The state after measurement $\rho_k$ and the probability $p_k$ are given by
\begin{equation}
    \rho_k=\frac{1}{p_k}(I\otimes B_k)\rho(I\otimes B_k),
\end{equation}
where 
\begin{equation}
    p_k=Tr[\rho(I\otimes B_k)],
\end{equation}
with
\begin{equation}\label{bk}
    B_k=V\Pi_kV^{\dagger}
\end{equation}
and
\begin{equation}
    \Pi_k=\ket{k}\bra{k}
\end{equation}
denoting a projector in the computational basis, so $k=0,1$, and $V \in SU(2)$. Notice we are here restricting ourselves to projective orthogonal measurements. The measurement-based mutual information $J(\rho|\{B_k\})$ is then 
\begin{equation}
    J(\rho|\{B_k\})=S(\rho_A)-S(\rho|{B_k}).
\end{equation}
with the measurement-based conditional entropy reading $S(\rho|{B_k}) = \sum_k p_k S(\rho_k)$. By optimizing over the least disturbing measurement basis, we define the classical correlation $C(\rho)$ as~\cite{Zurek,Henderson2001}
\begin{equation}
\begin{aligned}
       C(\rho)&=\max_{\{B_k\}}J(\rho|\{B_k\})\\
              &=S(\rho_A)-\min_{\{B_k\}}\sum_k p_kS(\rho_k) .
\end{aligned}
\end{equation}
Oppositely to the classical case, the quantities 
$I_m(\rho)$ and $C(\rho)$ are generally distinct in the quantum scenario. The difference between $I_m(\rho)$ and $C(\rho)$ is the QD 
\begin{equation}
     Q(\rho)=I_m(\rho)-C(\rho).
\end{equation}
In order to explicitly compute $Q(\rho)$, we can parametrize the unitary $V$, from Eq.~(\ref{bk}), as
\begin{equation}
V=\ket{\Psi}\bra{\Psi},
\end{equation}
where $\ket{\Psi}=\cos{(\frac{\theta}{2})}\ket{0}+\sin{(\frac{\theta}{2})}e^{i\phi}\ket{1}$. QD can then be explicitly obtained by performing an optimization over the angles $\theta$ and $\phi$~\cite{Girolami2011}.

%\textcolor{red}{ MDMS}\\

In order to build a MDMS, we have to maximize the amount of QD when compared with its classical correlation. A suitable strategy in this direction is to sacrifice some amount of entanglement to optimize QD. As show in Ref.~\cite{GALVEMSDS}, this can be achieved by the family of states
\begin{equation}
\begin{aligned}
    \rho^{MDMS}=&\epsilon\ket{\Phi^+(p)}\bra{\Phi^+(p)}+(1-\epsilon)[m\ket{01}\bra{01}\\&+(1-m)\ket{10}\bra{10}],
\end{aligned}
\label{rhomdms}
\end{equation}
where
\begin{equation}
    \ket{\Phi^+(p)}=\sqrt{p}\ket{00}+\sqrt{1-p}\ket{11}
    \label{PE}
\end{equation}
is a partial (unbalanced) Bell state. For $p=\frac{1}{2}$, we recover a maximally entangled Bell state. If we set the value $m=1$, Eq.~(\ref{rhomdms}) is reduced to a $(p,\epsilon)-$family of rank-2 states
\begin{equation}
    \rho^{(R2)}=\epsilon\ket{\Phi^+(p)}\bra{\Phi^+(p)}+(1-\epsilon)\ket{01}\bra{01}.
    \label{R2}
\end{equation}
On the other hand, for $p=\frac{1}{2}$, Eq.~(\ref{rhomdms}) yields a $(m,\epsilon)-$family of rank-3 states
\begin{equation}
\begin{aligned}
        \rho^{(R3)}=&\epsilon\ket{\Phi^+}\bra{\Phi^+}+(1-\epsilon)[m\ket{01}\bra{01}+\\&+(1-m)\ket{10}\bra{10}].
\end{aligned}
\label{R3}
\end{equation}
Eq.~(\ref{R3}) provides rank-3 states for different choices of parameters. A first rank-3 subset is $m \in [0,1]$ and $\epsilon \in [0,\frac{1}{3}]$. A second rank-3 subset is given by $m=1/2$ and $\epsilon \in [1/3,0.385]$. Eq.~(\ref{R3}) can also provide rank-2 states by taking $\epsilon$ in the interval $\epsilon \in [0.408,1]$.

\section{\label{sec:3}Spin-orbit maximally discordant mixed states}

A spin-orbit mode is described by polarization and transverse mode DoF. The polarization of light is defined as the oscillation direction of electric magnetic field. The circular polarization ($\sigma_{\pm}$) is associated with the intrinsic angular momentum of the photon (spin). The transverse modes are solutions of Helmholtz's paraxial equation and are responsible, for example, by the transverse shape of a laser beam. For a Cartesian coordinate system, the solutions are the well-known Hermite-Gauss ($HG_{m,n}(x,y)$) modes. For a cylindrical coordinate system, the solutions are the Laguerre-Gauss ($LG_{p}^l(\rho, \phi)$) modes, which can present orbital angular momentum (OAM)~\cite{siegman1986lasers}. Therefore, a transverse DoF of light is associated to orbital angular momentum. In the first-order subspace, $LG_0^{\pm1}$ can be written as a linear combination of $HG_{01}$ and $HG_{10}$~\cite{Padgett:99}. 
Given the linear polarization basis $\{\hat{e_H},\hat{e_V}\}$ and the first-order HG modes $\{HG_{01}(x,y),HG_{10}(x,y)\}$, the most general first order spin-orbit mode can be written as \cite{topo}
\begin{equation}
\begin{aligned}
    \vec{E_{SO}}(\vec{r})=& c_1HG_{10}(x,y)\hat{e}_H  +  c_2HG_{10}(x,y) \hat{e}_V  \\& +  c_3HG_{01}(x,y)\hat{e}_H
+ c_4 HG_{01}(x,y)\hat{e}_V, 
\end{aligned}
\label{spinorbitGEN}
\end{equation}
%\begin{eqnarray}
%{E_{SO}}(\vec{r})= c_1HG_{10}(\vec{r})\hat{e}_H  +  c_2HG_{10}(\vec{r}) \hat{e}_V  +  c_3HG_{01}(\vec{r})\hat{e}_H \nonumber \\ 
%+ c_4 HG_{01}(\vec{r})\hat{e}_V,
%label{spinorbitGEN}
%\end{eqnarray}
%
where $c_i$ are the complex numbers, with $i=1,2,3,4$. Such a basis can be used for the quantization of the electromagnetic field. Denoting $\hat{e_H}\equiv H$, $\hat{e_V}\equiv V$, $HG_{01}(\vec{r})\equiv h$, and  $HG_{10}(\vec{r})\equiv v$, the general quantum state of a photon can be written as 
\begin{equation}
\begin{aligned}
    \ket{\psi_{SO}}=& a_{Hh} \ket{Hh}  +  a_{Hv} \ket{Hv} + a_{Vh} \ket{Vh} \\& +  a_{Vv} \ket{Vv},
\end{aligned}
\label{spinorbitQSTATE}
\end{equation}
where $a_{ij}$ is the probability amplitude of the normalized basis element $\ket{ij}$, with $i=H,V$ and $j=h,v$. Eq. (\ref{spinorbitQSTATE}) represents a  general two-qubit state in DoF of polarization and transverse modes, with the computational basis defined as $\{ \ket{0}=\ket{H}, \ket{1}=\ket{V} \}$ and $\{ \ket{0}=\ket{h}, \ket{1}=\ket{v} \}$, respectively. 
For $a_{Hh}=a_{Vv}=1/\sqrt{2}$ and  $a_{Hv}=a_{Vh}=0$, we have 
\begin{equation}
\ket{\Phi^+_{SO}}= \frac{1}{\sqrt{2}} ( \ket{Hh}  +  \ket{Vv}),
\label{spinorbitBELL}
\end{equation}
which is a Bell state with maximal entanglement between internal DoF of a single photon (spin-orbit Bell state)~\cite{topo}. For $a_{Hv}=a_{Vh}=0$ and different arbitrary real $a_{Hh}$ and $a_{Vv}$, we have a spin-orbit partially entangled state, as described in Eq.(\ref{PE}). To produce $\ket{\Phi^+(p)}$, which is an essential ingredient for preparation of $\rho^{MDMS}$, we need to coherently superpose the single photon states $\ket{Hh}$ and $\ket{Vv}$ with different weights. It is important to stress that a spin-orbit product state  $\ket{Hv}=\ket{01}$ and $\ket{Vh}=\ket{10}$ [the needed states for the second part of the MDMS on the r.h.s of Eq.~(\ref{rhomdms})] are not difficult to prepare. Then, if we have these three ingredients, we can perform an incoherent superposition controlling the population of each one in order to produce $\rho^{MDMS}$. We can obtain the incoherent superposition by using three independent single-photon source to produce each one of the three required states. 

\begin{figure}[ht]
    \centering
    \includegraphics[width=0.49\textwidth]{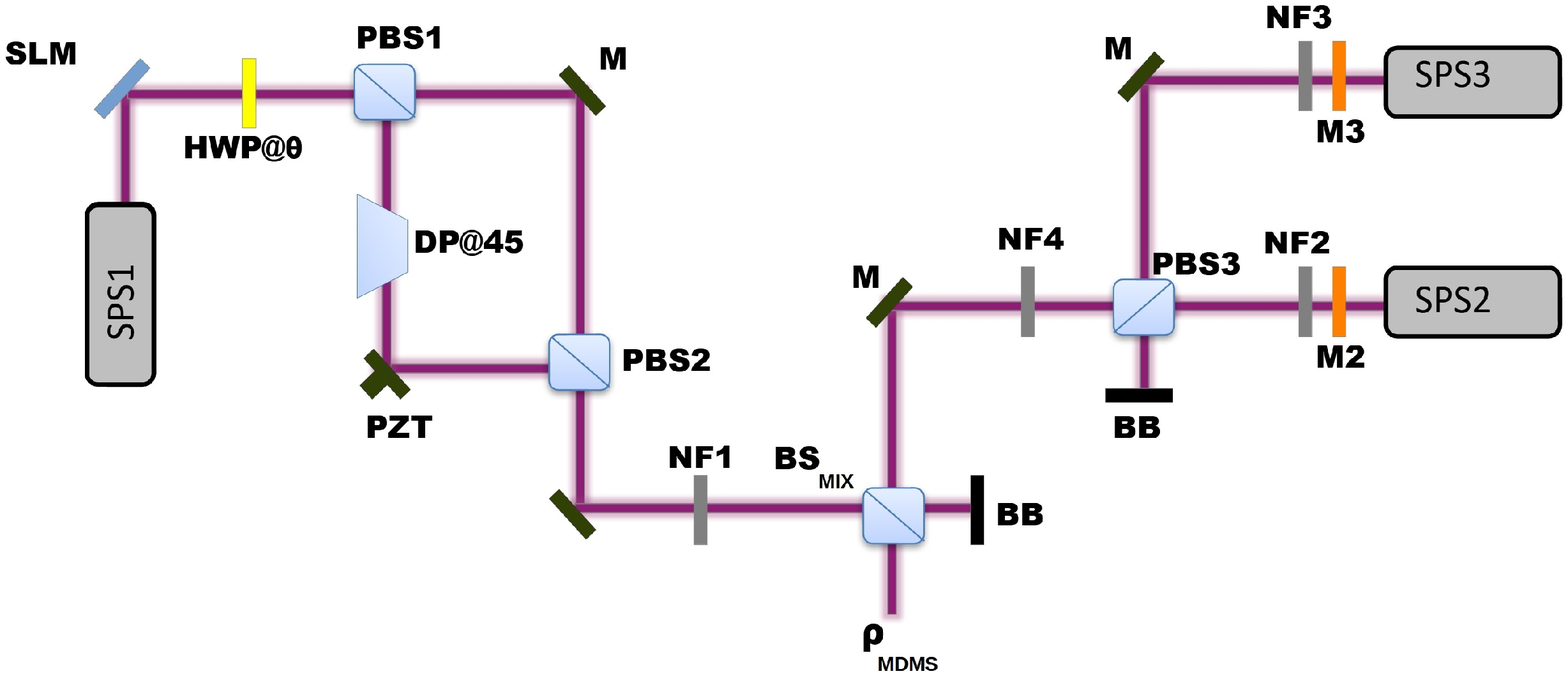}
    \caption{Proposed optical circuit. SPS stands for single photon source, SLM for spatial light modulator,  HWP@$\theta$ for half-wave plates with fast axis performing an angle $\theta$ with the horizontal, PBS for polarizer beam splitters, DP@$45$ for Dove prism rotated by $45^\circ$, PZT for Piezoelectric transducer  , NF for neutral filter, M2 and M3 for holographic masks for preparation of the $HG_{0,1}$ and $HG_{1,0}$ modes, and M for mirrors.}
    \label{circuit}
\end{figure}

We propose in Fig.\ref{circuit} a linear optical circuit to produce $\rho^{MDMS}$. Let us start with a partial entangled state. A first single-photon source (SPS1) prepares a photon horizontally polarized ($\ket{H}$) with probability $p^I$\footnote{hyper-written index is used for probability of photon production by a SPS}. A spatial light modulator (SLM) is used to prepare a HG transverse mode ($\ket{h}$). The photon state is then $\ket{Hh}$. A half-wave plate with an angle $\theta$ with respect the horizontal (HWP1$@\theta$) produces the superposition 
\begin{equation}
    \ket{Hh} \longrightarrow cos(2\theta)\ket{Hh} + sin(2\theta)\ket{Vh}.
\end{equation}
This superposition enters in a March-Zhender (MZ)-like interferometer mounted with two Polarizer Beam Splitters (PBS) and two mirrors. The arm of the reflected incident beam in PBS1 (vertical polarization) has a Dove prism rotated by $45^\circ$ with respect to the horizontal direction (DP$@45^\circ$) in order to transform $\ket{h}\rightarrow \ket{v}$. The mirror of this arm is mounted in a piezoeletric ceramic transdutor (PZT) to control the phase difference with respect to the other arm of the interferometer. Both arms are recombined in PBS2 and we have the state
\begin{equation}
    \ket{\Phi(\theta)} = cos(2\theta)\ket{Hh} + e^{i\phi}sin(2\theta)\ket{Vv}, 
\end{equation}
where $\phi$ is the difference of phase in the MZ interferometer. By setting $\phi=2n\pi$, $n$ as an integer, and $\sqrt{p}=cos(2\theta)$, we recover the partial entangled state of Eq.(\ref{PE}). Note that if the SLM prepares the transverse photon state $\ket{v}$, we have a possibility to build an odd partial entangled state ($\ket{\Psi^\pm}= cos(2\theta)\ket{Hv}\pm sin(2\theta)\ket{Vh}$). The signal $\pm$ is set by $\phi$ ($2n\pi$ for $+$ and $(2n+1)\pi$ for $-$ ).

The output of the interferometer passes through a neutral filter (NF1) with amplitude transmission $t_1=\sqrt{\epsilon}$ and will control the population of the state $\ket{\Phi^+(\theta)}$ in the spin-orbit DoF for $\rho^{MDMS}$. After NF1, the photon arrives at a balanced beam splitter (BS) responsible for the incoherent mixture (BS$_{MIX}$) that will transmit or reflect the photon with probability $1/2$. The transmitted output of the BS$_{MIX}$ is blocked (BB) and only the reflected output is used. In the output of BS$_{MIX}$ we have $\epsilon\ket{\Phi^+(\theta)}\bra{\Phi^+(\theta)}$ with probability $p^I/2$.

%For Rank-2 states given by Eq.(\ref{R2}), we need to use a second independent single-photon source (SPS2) that produce a  horizontally polarized photon state($\ket{H}$) with probability $p_2$ in order to prepare the product state $ket{Hv}=\ket{01}$. For this task, we can use a fixed holographic mask  M2 \cite{holomask} to produce a $\ket{v}$ transverse mode state. Then, the photon passes trough a Neutral Filter (NF2) adjusted for total transmission ($t_2=1$)  and is also transmitted with probability $1$ by the PBS3. The photon finds a new Neutral Filter (NF4) with amplitude transmission $t_4=\sqrt{(1-\epsilon)}$ that will control the population of the product state in $\rho^{MDMS}$ and, finally, arrives in the the orthogonal face BS$_{MIX}$ compared with the arriving face of the photon produced by SPS1. This photon will be reflected (and blocked) with probability $1/2$ and be transmitted in the same direction of the photon with state $\ket{\Psi^+(\theta)}$ also with probability $1/2$. After these transformation we have the spin-orbit Rank-2 state 
%
%\begin{equation}
 %   \rho^{(R2)}_{SO} = \epsilon \ket{\Psi^+(\theta)}\bra{\Psi^+(\theta)}  + (1-\epsilon)\ket{Hv}\bra{Hv},
 %   \label{R2SO}
%\end{equation}
%
%with probability $p_1p_2/4$. 

%Let us now present the preparation of Rank-3 states given by Eq.(\ref{R3}). 

To obtain the part of the product state in $\rho^{MDMS}$, we need to use a second independent single-photon source (SPS2) that produces a  horizontally polarized photon state ($\ket{H}$) with probability $p^{II}$. This will be able to yield the product state $\ket{Hv}=\ket{01}$. We also need a third independent single-photon source (SPS3) that produces a photon vertically polarized ($\ket{V}$) with probability $p^{III}$. The holographic masks M2 and M3 will produce the transverse mode photon state $\ket{v}$ and $\ket{h}$, respectively. Then SPS2 (SPS3) will give a rise the state $\ket{Hv}=\ket{01}$ ($\ket{Vh}=\ket{10}$).  The photon prepared by SPS2 passes trough neutral filter NF2 that has an amplitude transmission $t_2=\sqrt{m}$ and will be transmitted by PBS3 with probability $1$. On the other hand,  the photon prepared by SPS3 passes trough neutral filter NF3 that has an amplitude transmission $t_3=\sqrt{1-m}$ and will be reflected in PBS3 with probability $1$. Then, both photons pass through neutral filter NF4 that is adjusted with amplitude transmission $t_4=\sqrt{1-\epsilon}$. After NF4 these photons are sent to BB$_{MIX}$ and will be transmitted with probability $1/2$ each one. Then, from the contribution of the three SPS we will have
\begin{equation}
\begin{aligned}
           \rho^{MDMS}_{SO} =& \epsilon \ket{\Phi^+(\theta)}\bra{\Phi^+(\theta)}  + (1-\epsilon)[ m\ket{Hv}\bra{Hv} +\\&+ (1-m)\ket{Vh}\bra{Vh},
\end{aligned}
\label{Rhomdms}
\end{equation}
with probability $p^{I}p^{II}p^{III}/8$. 

In order to prepare a rank-2 state with $p=1/2$, we need to set $\theta=22.5^\circ$ and $m=1$, leading to $t_2=1$ and $t_3=0$, yielding
\begin{equation}
  \rho^{(R2)}_{SO} = \epsilon \ket{\Phi^+_{SO}}\bra{\Phi^+_{SO}}  + (1-\epsilon)\ket{Hv}\bra{Hv},
   \label{R2SO}
\end{equation}
where $\ket{\Phi^+_{SO}}$ is given by Eq.(\ref{spinorbitBELL}).

In order to prepare a rank-3 state with $p=1/2$, we need to set $\theta=22.5^\circ$ to obtain
\begin{equation}
\begin{aligned}
    \rho^{R3}_{SO} =& \epsilon \ket{\Phi^+_{SO}}\bra{\Phi^+_{SO}}  + (1-\epsilon)[ m\ket{Hv}\bra{Hv} +\\& + (1-m)\ket{Vh}\bra{Vh}].           
\end{aligned}
\label{R3SO}
\end{equation}
In the next section we will show the results for the simulation of these circuits and the QD calculations for each class of states.

\section{\label{sec:4-conclusions}Simulations of optical circuits}

The MDMS preparation was proposed to be implemented through a linear optical circuit. In this case, the action of the optical elements composing the circuit over the spin-orbit  photon state can be modeled by Jones matrix formalism \cite{ClarkJones:42}. For polarization states, the Jones matrices for the optical elements, such as HWP, PBS, and BS, are well known. For transverse modes, this is a non-trivial task. However, as we are limited to the first-order transverse modes ($HG_{01}$ and $HG_{10}$ modes), the analogy of this basis with linear polarization basis $\{ \hat{e_H},\hat{e_V} \}$ is straightforward. Indeed, we can construct a Poincaré-like sphere for first-order transverse modes, where first-order Laguerre-Gaussian beam plays the role of circular polarization \cite{Padgett:99}. Recently this powerful geometrical representation was extended for higher order modes \cite{PhysRevA.102.031501}. Then, we can represent a transverse mode state as a Jones vector and write unitary matrices for optical elements acting in the transverse mode. For example, the Jones matrix for a Dove prism acting on first order HG modes are analogue to HWP Jones Matrix acting in $\{ \hat{e_H},\hat{e_V} \}$ basis \cite{Xstate}.

 \begin{figure}[ht]
    \centering
    \includegraphics[scale=0.5,trim=0cm 0cm 3cm 0cm, clip=true]{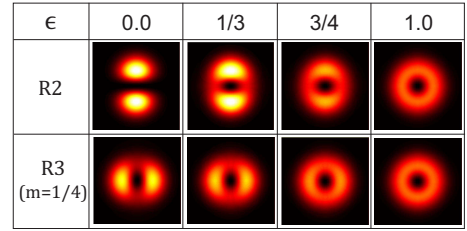}
    \caption{Density probability of detection of photon in the outuput transverse plane of the preparation circuit for different wheigh $\epsilon$. The calculation where performed for Rank-2 (R2) and Rank-3 (R3) states. We used $m=1/4$ for Rank-3 state}.
    \label{densityprob}
    \end{figure} 

By using Jones formalism, we construct a simulation of the optical circuit of Fig.\ref{circuit} with MATLAB\textregistered. Each Jones matrix for an optical element acts on the photon spin-orbit state accordingly. The resulting spin-orbit state is characterized by a two-qubit tomography for spin-orbit DoF, which is described in Ref.~\cite{Xstate}. Once we obtain the density matrix, we can calculate the classical correlation ($C$), pairwise entanglement as measured by concurrence ($C^\prime$)~\cite{Wootters:98}, and QD ($Q$) as shown in Section II. 

Let us now present the simulation results.  
Fig.~\ref{densityprob} exhibits the transverse density probability of photon detection that is related with the transverse profile the photon was prepared. For rank-2 (R2) states, we have pure state for $\epsilon=0$ (product state) and  $\epsilon=1$ (spin-orbit entangled state). For intermediate $\epsilon$ values we have a mixed state with the mixing of the density probabilities. For rank-3 states (R3), we set $m=1/4$. In this case, for $\epsilon=0$ we have a mixed state of the two product states $\ket{Hv}$ and $\ket{Vh}$. Then, we have the prevalence of the $\ket{Vh}$ mode, given the chosen $m$. For $\epsilon=1$, we also have a spin-orbit entangled state. 
 
\begin{figure}[h]
    \centering
    \includegraphics[width=0.48\textwidth]{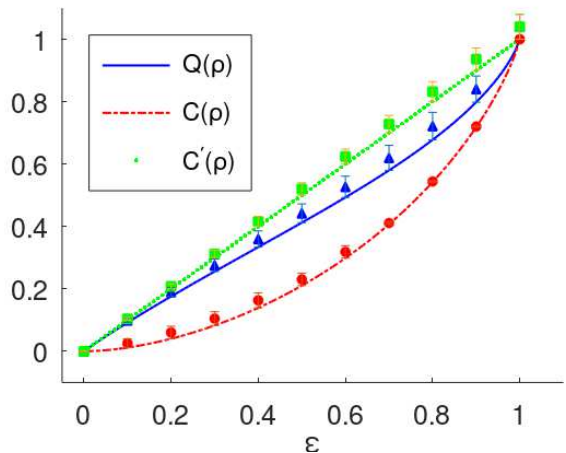}
    \caption{Correlations for rank-2 states ($p=0.5$) as function of the weigh $\epsilon$. Dots are the results for the calculation from the state simulated by the preparation optical circuit using Jones formalism. The lines are theoretical prediction of quantum theory.  Classical correlation $C$ : circles dots and dashed line (red online). Concurrence $C^\prime$: squares dots and dotted line (green online). Quantum discord $Q$: triangles dots and solid line (blue online).}
    \label{ResRank2}
    \end{figure}

The results for the correlations evaluated for R2 sates are presented in Fig.~\ref{ResRank2} by setting $p=0.5$. The circles dots (red online) are the results for classical correlation $C$ calculated as a function of $\epsilon$.  The dashed line is the theoretical prediction for classical correlation $C$ calculated as a function of $\epsilon$.  The squares (green online) are the results for concurrence $C^\prime$ as a function of $\epsilon$ and the dotted line is the prediction of the concurrence by quantum theory. Finally, triangles dots (blue online) are the results of quantum discord (Q) as a function of $\epsilon$ and the solid line is the discord predicted by quantum theory. 
The error bars were obtained by we included typical realistic sources of noise of linear optical circuits through optical components such as Half Wave Plate angles ($ \pm 1^\circ$), transmission and reflection in Beam Splitters of tomographic circuit ($R=48\%, T=49\%$). For this, we performed simulation varying the parameters in this error range and performed a statistical analysis of the results for different circuit runs, each with different sensitive parameters. As it can be seen, the error predicted for typical noise source in optical devices is very small. In addition, it is worth to stress the very good agreement between the results calculated from the simulated state by the preparation circuit with the predictions of quantum theory for MDMS.

For $\epsilon=0$ we have a product state and all correlations are vanishing, as expected. On the other hand, for $\epsilon=1$, we have a maximally entangled Bell state and all correlations are equal to $1$. Notice that all the correlations increase with $\epsilon$ being the concurrence dominant over QD and classical correlation.

\begin{figure}[ht]
    \centering
    \includegraphics[width=0.5\textwidth]{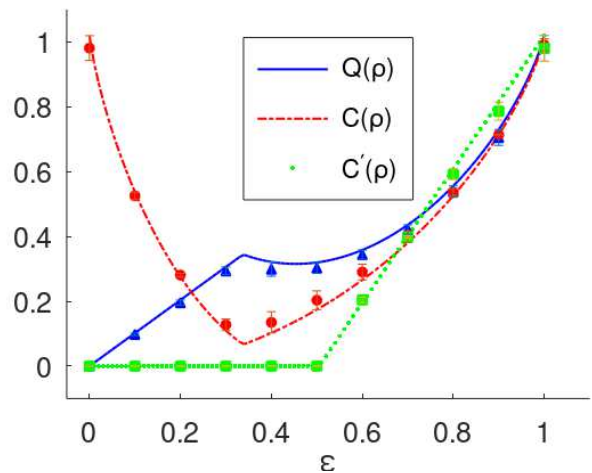}
    \caption{Correlations for rank-3 states ($p=m=0.5$) as function of the weigh $\epsilon$. Dots are the results for the calculation from the state simulated by the preparation optical circuit using Jones formalism. The lines are theoretical prediction of quantum theory.  Classical correlation $C$ : circles dots and dashed line (red online). Concurrence $C^\prime$: squares dots and dotted line (green online). Quantum discord $Q$: triangles dots and solid line (blue online).}
    \label{ResRank3}
    \end{figure}

The results for the R3 states are presented in Fig.~\ref{ResRank3} for $p=m=0.5$. For this case, we have a mixing of a maximally entangled spin-orbit state and a balanced $\ket{Hv}$ and $\ket{Vh}$ product mixed state.  As we can see, for $\epsilon$ from $0$ to $0.5$, the concurrence is vanishing but not QD. It is worth mentioning that, for $1/3 \leq \epsilon \leq 0.385$, we have found the minimal value for classical correlation and the local maximum for QD, which is the characteristic property of the MDMS. Again, the predicted error form realistic optical devices are small and the agreement between the results for discord calculated from the simulated state by the preparation circuit with the predictions of quantum theory for MDMS is remarkable. 

\begin{figure}[ht]
    \centering
    \includegraphics[width=0.50\textwidth]{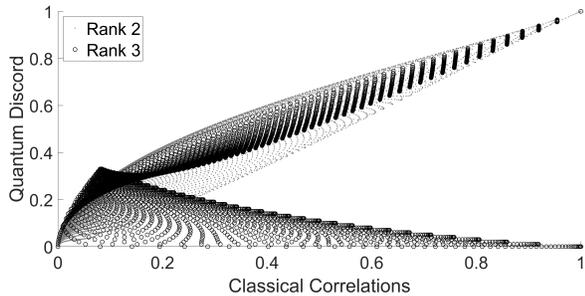}
    \caption{Quantum Discord versus Classical Correlation for a sample of rank-2 (gray dots) and rank-3 (black circles) states.}
    \label{ResGeral}
    \end{figure}

A global analysis can be performed by looking for QD as a function of classical correlation, as shown in Fig.\ref{ResGeral}. The results were obtained by varying $p$ in steps of $0.01$. Gray dots are the results for R2-states [Eq.~(\ref{Rhomdms}) for $m=0$] and the black circles for the R3-states [Eq.~(\ref{Rhomdms}) for $m=0$]. For a low value of the classical correlation, the QD for R3 states ($\theta=22.5^\circ \rightarrow p=1/2$, upper bound of black circles) is higher than the QD for R2 states ($\theta=22.5^\circ \rightarrow p=1/2$, upper bound of gray dots). On the other hand, when classical correlation increases, R2 states presents higher QD than R3 states. This result shows excellent agreement  with the theoretical predictions presented in Ref.~\cite{GALVEMSDS}.  

\section{\label{sec:5-conclusions}Conclusions}

We have proposed an optical circuit to produce  classes of rank-2 and rank-3 spin-orbit mixed states as well as others classes of mixed states, including MDMS. This optical circuit provides a useful tool to probe states with optimized quantum correlations for a fixed amount of classical correlation. Remarkably, for the case of rank-3 states, the circuit allows to explore the optimization of quantum correlations (as measured by QD) in the absence of pairwise entanglement. The circuit has been simulated in a realistic experimental scenario, with the theoretical and simulated correlations showing excellent agreement. As a further development, we expect to provide an experimental realization of the circuit. This will allow for the explicit investigation (and control in certain cases) of decoherence as well as its impact on each kind of correlation. This is left for a future work. 

\begin{acknowledgments}

We would like to thank financial support from Conselho Nacional de Desenvolvimento Cient\'{\i}fico e Tecnol\'ogico (CNPq), Funda\c{c}\~ao Carlos Chagas Filho de Amparo \`a Pesquisa do Estado do Rio de Janeiro (FAPERJ), Coordena\c{c}\~ao de Aperfei\c{c}oamento de Pessoal de N\'{\i}vel Superior  - Brasil (CAPES) (Finance Code 001), the Brazilian National Institute for Science and Technology of Quantum Information (CNPq INCT-IQ), and W.F.B has financial support from the grant H2020-FETOPEN Grant PHOQUSING ((GA no.: 899544). 

\end{acknowledgments}

\bibliography{citations}% Produces the bibliography via BibTeX.

\end{document}